\documentstyle[12pt]{article}
\def\beq{\begin{equation}}
\def\eeq{\end{equation}}
\def\bea{\begin{eqnarray}}
\def\eea{\end{eqnarray}}
\def\nn{\nonumber}
\def\tp{\tilde P}
\def\td{\tilde D}
\def\ckr{\check R}
\begin{document}
\begin{tabbing}
\` OS-GE-34-94 \\
\` RCNP-058 \\
\` August 1993
\end{tabbing}
\begin{center}
\vfill
{\large Quantum Affine Transformation Group and Covariant
Differential Calculus}

\vspace{2cm}

N. Aizawa$^*$
\vspace{0.5cm}

{\em Research Center for Nuclear Physics}

{\em Osaka University, Ibaraki, Osaka 567, Japan}

\vspace{0.7cm}
and

\vspace{0.7cm}
H.-T. Sato$^{\dagger}$
\vspace{0.5cm}

{\em Institute of Physics, College of General Education}

{\em Osaka University, Toyonaka, Osaka 560, Japan}

\end{center}

\vfill
\begin{abstract}
  We discuss quantum deformation of the affine transformation group and
its Lie algebra. It is shown that the quantum algebra has a non-cocommutative
Hopf algebra structure, simple realizations and quantum tensor operators.
The deformation of the group is achieved by using the adjoint representation.
The elements of quantum matrix form a Hopf algebra.
Furthermore, we construct a
differential calculus which is covariant with respect to the action of
the quantum matrix.
\end{abstract}


\noindent-------------------------------------------------------------------\\
$^{\dagger}$ {\footnotesize Fellow of the Japan Society for the
Promotion of Science} \\
\mbox{}\hspace{0.3cm}{\footnotesize E-mail~:
hsato@jpnyitp.yukawa.kyoto-u.ac.jp} \\
$^*$ {\footnotesize E-mail~: aizawa@rcnpth.rcnp.osaka-u.ac.jp}
\newpage
%
%
\section{Introduction}

  Recent progress in understanding quantum integrability and non-commutative
geometry has introduced the notion of quantum deformation of groups and
Lie algebras [1 - 6]. The quantum deformation of
groups is characterized by the fact that the elements of its representation
matrix do not mutually commute. And the quantum deformation of  Lie
algebras is understood that its universal enveloping algebra has the
structure of non-cocommutative Hopf algebra. The former is called the quantum
group and the latter the quantum algebra.

 The paradigm of the quantum group is $ SL_q(2) $, which is the deformation
of $ 2 \times 2 $ matrix group
and the matrix elements obey certain commutation relations
depending on a deformation parameter $ q$ \cite{ma}. The algebra dual to
$ SL_q(2) $ is the quantum algebra $ U_q(sl(2)) $\cite{tak,maj}. Various
aspects of $ SL_q(2) $ and $ U_q(sl(2)) $ have been investigated by many
authors. The quantum deformation of other groups and algebras of classical,
exceptional and super has also been discussed. Furthermore, some
generalizations to multi-parameter deformation have been attempted
\cite{mul}.

  Standing point of view of the theoretical physicists, the physical phenomena
for
which quantum group would be available are interesting. Continuous
transformations of physical quantities are elements of continuous groups,
and consequently, the groups play fundamental roles in physics. For example,
the spatial and time translation forms the Abelian group and the invariance
under their operations imply the conservation laws of momentum and energy.
 In field theories, local gauge invariance uniquely determines
the interactions between matters and gauge fields. Because the transformation
law is described by a representation of the group considered in the system,
the representation of quantum group is also an issue of wide importance to
physical application accordingly.

  In this paper, we examine the quantum deformation of the group and the
algebra
of the affine transformation. The generators of the affine transformation
group are provided by
\beq
  p = -i {d \over dx}, \hspace{1cm}
  d = {i \over 2}(x{d \over dx} + {d \over dx} x),     \label{eq11}
\eeq
where $ p $ is the momentum operator in one-dimensional quantum mechanics
and $ d $ is the dilatation operator. The affine transformation group
consists of two physically important
operations ;
\beq
 \exp{(-iap)}\, x \,\exp{(iap)} = x - a,               \label{eq12}
\eeq
and
\beq
 \exp{(i\ln b\, d)}\, x \,\exp{(-i\ln b\, d)} = bx.           \label{eq13}
\eeq
Hence it is interesting and important to investigate the quantum deformation
of the affine transformation group in the abovementioned sense.
The aim of this paper is 1) to show
that one-parameter deformation of both the affine transformation group
and the
algebra can be established as non-commutative and non-cocommutative Hopf
algebras, 2) to develop the representation theory of the quantum affine
transformation algebra (QATA), 3) to construct differential calculus on the
quantum plane \cite{wz} which is covariant under the action of the quantum
affine transformation group (QATG).

  This paper is organized as follows. In the next section, we briefly review
the representation of the affine transformation group and the Lie algebra.
In \S 3, we discuss the QATA. Some realizations of the QATA and a quantum
analogue of the adjoint representation are constructed. The quantum analogue
of the tensor operator \cite{ten} which carries the adjoint representation
is explicitly given in terms of the generators of the QATA. In \S 4, we
establish the the QATG on the bases of adjoint representation. The matrix
elements
of the QATG satisfy the same commutation relation as that of Manin's
quantum plane \cite{ma}, however it leads to new R-matrices, {\em i.e.}
new solutions of
the Yang-Baxter equation (YBE). \S 5 is devoted to a differential calculus
on the quantum plane which is covariant under the action of the the QATG.
It is shown that the covariance requires a somewhat different quantum plane
from
that of Manin \cite{ma}. Finally, conclusion and discussion are in \S 6.
%
%
\setcounter{equation}{0}
\section{Affine Transformation Group}

  In this section, we give a short review of the affine transformation
group. The affine transformation group is abstractly defined as the group
of linear transformations without reflection on real line :
$ x \rightarrow bx - a $. In our particular parametrization the elements of
the group are denoted by
\beq
  U(a,b) = \exp{(iap)}\,\exp{(-i\ln b\, d)},              \label{eq21}
\eeq
where the Lie algebra $ A = \{p, d\} $ satisfies the commutation relation
\beq
  [d,\, p] = ip.                                          \label{eq22}
\eeq
The allowed region of the parameters is
$ -\infty < a < \infty, \;\; 0 < b < \infty $,
which requires that the
group manifold is a half-plane, and so the affine transformation group is
a non-Abelian and non-compact group.
The group multiplication law is given by,
\beq
  U(a,b)\, U(\alpha,\beta) = U(a+\alpha b, \beta b).         \label{eq23}
\eeq

  The unitary representations of the affine transformation group are
found by
Gel'fand and Naimark \cite{gn}. Aslaksen and
Klauder gave an alternative proof \cite{ak}.
It was shown in Refs.\cite{gn,ak} that there exist two and
only two unitary and inequivalent irreducible representations. One is the
case of positive eigenvalue of $p$ and the other the case of negative.
Namely, when the representation space $ V $ is taken to the space of all
functions $ \phi(k) \in L^2({\bf R}) $,
\[
   \int^{\infty}_{-\infty}\,|\phi(k)|^2 dk \;\; < \;\; \infty,
\]
$ V $ is the direct sum of two invariant subspaces under the action of
$ U(a,b) $ ;
$
  V = V_+ \oplus V_-,
$
where
\bea
  & V_+ \equiv \{ \phi(k) \; :\; \phi \in L^2({\bf R}),\; \phi(k) = 0
    \;\; {\rm for}\;\; k \leq 0 \}, \nn \\
  & V_- \equiv \{ \phi(k) \; :\; \phi \in L^2({\bf R}),\; \phi(k) = 0
    \;\; {\rm for}\;\; k \geq 0 \}.
                                                        \label{eq25}
\eea
For example, if we take the following realization of the Lie algebra
$ A $
\beq
   p = k, \;\;\; d = {i\over 2}(k\partial_k + \partial_k k),
                                                         \label{eq26}
\eeq
then the space $ V_+ $ consists of only $ \phi(k) $ \cite{ak,kla}
\[
  \phi(k) = N k^{1/2}e^{-k},
\]
where $N$ is the normalization factor.
The action of $ U(a,b) $ is given by ,
\[
  U(a,b)\,\phi(k) = \exp{(iak)}\,b^{1/2}\phi(b^{1/2}k).
\]

  In the matrix representation of the affine
transformation group, the adjoint representation is the simplest one.
The adjoint representation of the algebra $A$
\beq
   p = \left(
       \begin{array}{cc}
       0 & 0 \\
       -i & 0
       \end{array} \right),\hspace{1.5 cm}
   d = \left(
       \begin{array}{cc}
       0 & 0 \\
       0 & i
       \end{array} \right),                                \label{eq28}
\eeq
gives the adjoint representation of the affine transformation group by
substitution of (\ref{eq28}) into (\ref{eq21})
\beq
   U(a,b) = \left(
       \begin{array}{cc}
       1 & 0 \\
       a & b
       \end{array} \right).                                \label{eq29}
\eeq
The adjoint representation is not a unitary one, since it is a finite
dimensional representation and the affine transformation group is
non-compact.

  We explicitly write down the adjoint action for later convenience
\bea
  & \varphi(p) = U(a,b)\, p \, U^{-1}(a,b) = b p, \nn \\
  & \varphi(d) = U(a,b)\, d \, U^{-1}(a,b) = d + a p.         \label{eq210}
\eea
$ \varphi(p) $ and $ \varphi(d) $ also satisfy the commutation
relation (\ref{eq22}).
%
%
\setcounter{equation}{0}
\section{Quantum Affine Transformation Algebra}
\subsection{QATA and its realization}

  We present a one-parameter deformation of the universal enveloping algebra
of $ A $ {\em i.e.} the QATA and its realizations. First, we define
QATA as the algebra which is generated by two elements $ P $ and $ D $
satisfying the commutation relation
\beq
  [D,\, P] = i[P],                                       \label{eq31}
\eeq
where
$ [P] \equiv (q^P - q^{-P}) / (q - q^{-1}) $ and $ q $ is as usual the
deformation parameter. This is a non-cocommutative Hopf algebra. The Hopf
algebra mappings, coproduct $ \Delta $, counit $ \epsilon $ and
antipode $ S $, are given by
\bea
  & \Delta(P) = P \otimes 1 + 1 \otimes P,        \nn \\
  & \Delta(D) = D \otimes q^{-P} + q^P \otimes D, \nn \\
  & \epsilon\,(P) = \epsilon\,(D) = 0,            \label{eq32}    \\
  & S(P) = - P,\;\; S(D) = -D -i\, (\ln q) [P],   \nn
\eea
and eqs.(\ref{eq32}) certainly satisfy the following axioms of the
Hopf algebra
\bea
  & (id \otimes \Delta) \circ \Delta = (\Delta \otimes id) \circ \Delta,
  \nn \\
  & (id \otimes \epsilon) \circ \Delta = (\epsilon \otimes id) \circ \Delta
    = id,
  \label{eq33} \\
  & m(id \otimes S) \circ \Delta = m(S \otimes id) \circ \Delta = 1 \epsilon,
  \nn
\eea
where $ id $ denotes the identity mapping and $m$ the product of
the two terms in the tensor product ;
$ m(x \otimes y) = xy $. If we define the opposite coproduct $ \Delta'$ by
\beq
   \Delta' = \sigma \circ \Delta, \hspace{1cm}
   \sigma(x \otimes y) = y \otimes x,                   \label{eq34}
\eeq
$ \Delta'(P) $ and $ \Delta'(D) $ also satisfy the same commutation
relation as
(\ref{eq31}). For the opposite coproduct,
$ S(P) $ and the counit are not changed while
$ S(D) $ becomes
\beq
  S(D) = -D + i\,(\ln q) [P].                            \label{eq35}
\eeq

  Next, we show some realizations of the QATA. The generators $P$ and $D$
can be formally expressed in terms of the undeformed ones
\beq
  P = p, \hspace{1cm} D = {1 \over 2}({[p] \over p} d + d {[p] \over p}).
                                                         \label{eq36}
\eeq
When the representation of $p$ and $d$ in the Hilbert space is considered,
$p$ and $d$ are hermite operators. The realization (\ref{eq36})
of $ P $ and $ D $
is also chosen to be hermitian in the same representation space when
$ q $ is real or $ |q| = 1 $. If we require only satisfying the commutation
relation (\ref{eq31}), $ D $ can be simply given by
\beq
  D = {[p] \over p} d.                                        \label{eq37}
\eeq
When the representation and the realization of $A $ have the inverse
$ p^{-1} $ or $ [p] $ are proportional to $ p $, they can be transformed into
those of the QATA by making use of (\ref{eq36}) or (\ref{eq37}).

\vspace{0.7 cm}

\noindent
{\bf Examples}

\noindent
(1) The case of having $ p^{-1} $. The realization of eq.(\ref{eq26}) is
transformed into
\beq
  P = k, \hspace{1 cm} D = {i \over 2} ([k]\partial_k + \partial_k [k]),
                                                              \label{eq38}
\eeq
it is easy to verify that the commutation relation (\ref{eq31}) holds.

\noindent
(2) On the other hand, the adjoint representation of the QATA cannot be
obtained by naive use of the relation (\ref{eq36}) since the
adjoint representation of the undeformed generators (\ref{eq28}) does not have
$ p^{-1} $. However $ [p] $ reduces to be proportional to $ p $, {\em i.e.}
\[
  [p] = \delta p, \hspace{1 cm} \delta \equiv {2 \ln q \over q - q^{-1}},
\]
so we get the adjoint representation of $ P $ and $ D $ with the aid of
eq.(\ref{eq36}) ;
\beq
   P = \left(
       \begin{array}{cc}
       0 & 0 \\
       -i & 0
       \end{array} \right),\hspace{1.5 cm}
   D = \left(
       \begin{array}{cc}
       0 & 0 \\
       0 & i\delta
       \end{array} \right).                                \label{eq39}
\eeq
The adjoint representation of the QATA is the matrices (\ref{eq28}) multiplied
by $q$-dependent factor $ \delta $. The factor $ \delta $ becomes
unity as $ q \rightarrow 1 $.
\subsection{Tensor Operators}

  In this subsection, we show explicit expression of the tensor operator
which carries the adjoint representation of the QATA. The definition of tensor
operators of a quantum algebra was given by Rittenberg  and Scheunert
\cite{ten} in terms of the representation theory of the Hopf algebra.
The tensor operator is generally defined through the following adjoint
action of the Hopf algebra $H$.
The adjoint action of $ c \in H$ on $ t \in H$ is defined by
\beq
  ad(c)\, t = \sum_i c_i\, t\, S(c'_i),                         \label{eq310}
\eeq
where we denote the coproduct of $ c$ by
\[
  \Delta(c) = \sum_i c_i \otimes c'_i.
\]
Writting the $ n \times n $ matrix representation of $ c $ as
$ \rho_{ij}(c) $,
the tensor operators $ \{ T_i, \; i = 1,\,2,\, \cdots,\, n\} $ which
carry the representation $ \rho(c) $ are defined by the relation
\beq
   ad(c)\, T_i = \sum_j \rho_{ji} (c) T_j.                      \label{eq311}
\eeq
Namely, the tensor operators $ \{ T_i \} $ form a representation basis under
the adjoint action.

  Now in the case of the QATA, we can write down the adjoint action of
$ \{P,D \} $
on an element of the QATA $ T $
\bea
   & ad(P)\, T = [P,\, T], \nn \\
   & ad(D)\, T = D T q^{P} - q^P T D - i(\ln q) q^P T [P].      \label{eq312}
\eea
We therefore find the tensor operators which carry the adjoint representation
of (\ref{eq39}) :
\bea
   & T_1 = q^{-P} D,   \nn \\
   & T_2 = q^{-P} [P].                                          \label{eq313}
\eea
It is noted that RHS of eq.(\ref{eq312}) reduces to the commutators
and eq.(\ref{eq313}) to $ d $ and $ p $ in the limit of
$ q \rightarrow 1 $.
\subsection{Intertwiner and Yang-Baxter Equation}

  As is mentioned in \S 3.1, we have two coproduct ; $ \Delta $ and the
opposite
$ \Delta'$. In conformity with this fact, we have two tensor product
representations and are urged to question on an intertwiner between them. In
this
subsection, the intertwiner in the case of the adjoint
representations is investigated as an example. The coproduct of two adjoint
representations $ \Delta $ and the opposite are
$ 4 \times 4 $ matrices and we should find a $ 4 \times 4 $ matrix $ R $
which satisfies the relations of intertwiner
\beq
  R \Delta(P) R^{-1} = \Delta'(P), \hspace{1 cm}
  R \Delta(D) R^{-1} = \Delta'(D).
\label{eq314}
\eeq
The solution of (\ref{eq314}) is of the following form
\beq
  R =  \left(
       \begin{array}{cccc}
       r & 0 & 0 & 0 \\
       0 & r & g & 0 \\
       0 & 0 & r_{33} & 0 \\
       0 & r_{42} & r_{43} & r_{33}
       \end{array} \right)_,                                \label{eq315}
\eeq
where
\[
   r_{33} = {i - h \over i + h} r, \;\;\;
   r_{42} = {2 h \over i + h} r,
\]
\beq
   r_{43} = -ihg + 2i r_{33}, \;\;\;
   h = \ln q,
                                                            \label{eq316}
\eeq
The constants $ r $ and $ g$ are undetermined from eq.(\ref{eq314}).
We note that one
of the undetermined constants is meaningless because it is nothing but
the overall multiplication factor on $ R $.

  The next stage of our interests is whether the matrix $ R $ satisfies
the YBE
\beq
  R_{12}R_{13}R_{23} = R_{23}R_{13}R_{12},                  \label{eq317}
\eeq
where $ R_{12} $ {\em etc.} mean $ R \otimes id $ {\em etc.} which act on
the tensor product of three representation spaces.
Substituting (\ref{eq315}) into (\ref{eq317}), it is not difficult  to see
that the YBE is satisfied if and only if $ r = 0 $.
However, this case is not
legitimate since the matrix $ R $ dose not have the inverse.

  In summary, the intertwiner for the tensor product of two adjoint
representations exists, however it is incompatible with the YBE. From this
result, we deduce that the QATA dose not possess the universal R-matrix.
%
%
%
\setcounter{equation}{0}
\section{Quantum Affine Transformation Group}

  In this section, the quantum deformation of the affine transformation
group is discussed in the spirit of Manin \cite{ma}. The deformation of the
representation basis of the group induces that of the
representation matrix. This deformed matirx should be regarded as
the deformation of the group.

  We have a simple matrix representation of the affine transformation
group, {\em i.e.} the adjoint representation (\ref{eq29}). The
representation basis of the adjoint representation is the Lie algebra of
the affine transformation group (see eq.(\ref{eq21})). Hence
our construction of the QATG should be based on the deformation of
the Lie algebra.
The deformation which is considered in the previous section is not
appropriate for our purpose since RHS of.(\ref{eq31}) is an infinite
power series of $ P $ and so the deformed matrix becomes infinite
dimensional. Our aim here is to discuss the defomation of $ 2 \times 2 $ matrix
group.

  To this end, we use the following new basis instead of $ P $ and $ D $
\beq
   \td \tp - q \tp \td = i \tp,                               \label{eq41}
\eeq
where these new elements are transformed from those discussed in \S 3
as follows
\bea
  & & \tp = f(\Lambda) P,               \nn \\
  & & \td = iq^{(\Lambda - 1)/2} [\Lambda]_{1/2}.              \label{eq42}
\eea
$ f(\Lambda) $ is an arbitrary function of $ \Lambda $ provided that
$ f(\Lambda) \rightarrow 1 $ as $ q \rightarrow 1 $ and
\[
   \Lambda \equiv -i {P \over [P]} D,
\]

\beq
   [\Lambda]_{1/2} \equiv {q^{\Lambda /2} - q^{-\Lambda /2} \over
                           q^{1/2} - q^{-1/2}}.
                                                              \label{eq43}
\eeq

Let us consider the situation that the new basis $ \{ \td,\: \tp \} $
will become a comodule algebra of the adjoint representation matrix,
Namely, under the following transformation
\beq
  (\varphi(\td),\, \varphi(\tp)) = (\td,\, \tp) \,
  \left(
  \begin{array}{cc}
        1 & 0 \\
        a & b \\
  \end{array} \right)_,                                       \label{eq44}
\eeq
$ \varphi(\td) $ and $ \varphi(\tp) $ also
satisfy the relation (\ref{eq41}),
where we assume that $ \td $ and $ \tp $ commute with $ a $ and $ b $.
This requirement deforms $a$ and $ b$ into the non-commutative
objects which satisfy the `quantum plane' commutation relation
\beq
  a b = q b a.                                                \label{eq45}
\eeq
We thus define the QATG as the matrix
\beq
   \left(
   \begin{array}{cc}
   1 & 0 \\
   a & b
   \end{array} \right)_,                                     \label{eq46}
\eeq
whose elements satisfy the relation (\ref{eq45}).

  It is emphasized that eq.(\ref{eq46}) is not a `subgroup' of $ GL_q(2) $.
If it were so, it should preserve commutation relation $ x y = q y x $
of the quantum plane $ (x,\,y) $
under the action
\[
  \left( \begin{array}{c}
         x' \\
         y'
         \end{array} \right)
  = \left( \begin{array}{cc}
         1 & 0 \\ a & b
         \end{array} \right)
  \,
  \left( \begin{array}{c}
         x \\ y
         \end{array} \right),
\]
where $ (x,\, y) $ is assumed to commute with $ a $ and $ b $. Is is easy to
see that the commutation relation is not preserved by the action of the QATG.

  It is possible to express the defining relation of the QATG in terms of
the R-matrix. Considering the tensor product of (\ref{eq46}) and the $ 2 \times
2 $
unit matrix
\beq
  T_1 = \left( \begin{array}{cc}
               1 & 0 \\
               a & b
               \end{array} \right) \otimes {\bf 1},
  \hspace{1 cm}
  T_2 = {\bf 1} \otimes
               \left( \begin{array}{cc}
               1 & 0 \\
               a & b
               \end{array} \right)_,                       \label{eq47}
\eeq
the commutation relation (\ref{eq45}) should be expressed as
\beq
  R T_1 T_2 = T_2 T_1 R,                                   \label{eq48}
\eeq
where the $ 4 \times 4 $ matrix $ R $ is the R-matrix of the QATG
(we use the same notation as in \S 3.3,
but it will make no serious confusion).
The solutions of eq.(\ref{eq48}) are given by
\beq
  R^{(1)} = \left(
  \begin{array}{cccc}
  1 & 0 & 0 & 0 \\
  0 & q & 1-q & 0 \\
  0 & 0 & 1 & 0 \\
  0 & 0 & 0 & 1
  \end{array} \right)_,
  \hspace{1 cm}
  R^{(2)} = \left(
  \begin{array}{cccc}
  1 & 0 & 0 & 0 \\
  0 & 1 & 0 & 0 \\
  0 & 1-q^{-1} & q^{-1} & 0 \\
  0 & 0 & 0 & 1
  \end{array} \right)_,                                     \label{eq49}
\eeq
and it is verified that both $ R^{(1)} $ and $ R^{(2)} $ satisfy the YBE
\beq
    R^{(i)}_{12}\, R^{(i)}_{13}\, R^{(i)}_{23} =
    R^{(i)}_{23}\, R^{(i)}_{13}\, R^{(i)}_{12},
    \hspace{1 cm} i = 1,\, 2.                 \label{eq410}
\eeq
To make clear the relation between $ R^{(1)} $ and $ R^{(2)} $, we work with
\beq
   \ckr^{(i)} \equiv \sigma\, R^{(i)}, \hspace{1 cm}
   \sigma = \left(
   \begin{array}{cccc}
   1 & 0 & 0 & 0 \\
   0 & 0 & 1 & 0 \\
   0 & 1 & 0 & 0 \\
   0 & 0 & 0 & 1
   \end{array} \right)_,                                    \label{eq411}
\eeq
and the explicit formulae are given by
\beq
  \ckr^{(1)} = \left(
  \begin{array}{cccc}
   1 & 0 & 0 & 0 \\
   0 & 0 & 1 & 0 \\
   0 & q & 1-q & 0 \\
   0 & 0 & 0 & 1
  \end{array} \right)_,
  \hspace{1 cm}
  \ckr^{(2)} = \left(
  \begin{array}{cccc}
   1 & 0 & 0 & 0 \\
   0 & 1-q^{-1} & q^{-1} & 0 \\
   0 & 1 & 0 & 0 \\
   0 & 0 & 0 & 1
  \end{array} \right)_.                                   \label{eq412}
\eeq
{}From (\ref{eq412}), it turns out that $ \ckr^{(1)} $ is the inverse of
$ \ckr^{(2)} $. Both of the $ \ckr $-matrices satisfy the YBE
\beq
  \ckr^{(i)}_{12}\, \ckr^{(i)}_{23}\, \ckr^{(i)}_{12} =
  \ckr^{(i)}_{23}\, \ckr^{(i)}_{12}\, \ckr^{(i)}_{23},   \label{eq413}
\eeq
where
$ \ckr_{12} = \ckr \otimes {\bf 1} $ and
$ \ckr_{23} = {\bf 1} \otimes \ckr $.

  For later convenience, we write down the transposed matrix of (\ref{eq46}) as
well
\beq
   \left( \begin{array}{cc}
          1 & a \\
          0 & b
          \end{array} \right)_,                            \label{eq414}
\eeq
and define
\beq
   T_1' = \left( \begin{array}{cc}
          1 & a \\
          0 & b
          \end{array} \right) \otimes {\bf 1},
   \hspace{1 cm}
   T_2' = {\bf 1} \otimes
          \left( \begin{array}{cc}
          1 & a \\
          0 & b
          \end{array} \right).                               \label{eq415}
\eeq
The commutation relation (\ref{eq45}) can be expressed as
\beq
   R' T_1' T_2' = T_2' T_1' R',                               \label{eq416}
\eeq
with
\beq
  R'^{(1)} = \left(
  \begin{array}{cccc}
  1 & 0 & 0 & 0 \\
  0 & 1 & 1-q & 0 \\
  0 & 0 & q & 0 \\
  0 & 0 & 0 & 1
  \end{array} \right)_,
  \hspace{1 cm}
  R'^{(2)} = \left(
  \begin{array}{cccc}
  1 & 0 & 0 & 0 \\
  0 & q^{-1} & 0 & 0 \\
  0 & 1-q^{-1} & 1 & 0 \\
  0 & 0 & 0 & 1
  \end{array} \right)_.                                     \label{eq417}
\eeq
Again defining $ \ckr'^{(i)} \equiv \sigma\, R'^{(i)} $, it turns out that
$ \ckr'^{(1)} $ is the inverse of $ \ckr'^{(2)} $ and both
$ \ckr'^{(1)} $ and $ \ckr'^{(2)} $ satisfy the YBE (\ref{eq413}). The
relationship among $ \ckr^{(i)} $ and $ \ckr'^{(i)} $ is summarized in the
Figure~1.

  Next, we show that the algebra generated by $ a $ and $ b $ has only the
structure of bialgebra, however incorporating $ b^{-1} $ into the algebra,
they form a Hopf algebra. The coalgebra mappings of $ a $ and $ b $ are
given as follows.

  The coproduct is symbolized as
\beq
   \Delta \left(
   \begin{array}{cc}
   1 & 0 \\
   a & b
   \end{array} \right) =
   \left(
   \begin{array}{cc}
   1 & 0 \\
   a & b
   \end{array} \right) \otimes
   \left(
   \begin{array}{cc}
   1 & 0 \\
   a & b
   \end{array} \right)                                  \label{eq418}
\eeq
which is interpreted as
\beq
   \left(
   \begin{array}{cc}
   \Delta(1) & 0 \\
   \Delta(a) & \Delta(b)
   \end{array} \right)
   =
   \left(
   \begin{array}{cc}
   1 \otimes 1  & 0  \\
   a \otimes 1 + b \otimes a & b \otimes b
   \end{array} \right)_.                                   \label{eq419}
\eeq
In the same notation, the counit is given by
\beq
   \epsilon\, \left(
   \begin{array}{cc}
   1 & 0 \\
   a & b
   \end{array} \right)
   =
   \left(
   \begin{array}{cc}
   1 & 0 \\
   0 & 1
   \end{array} \right)_.                                     \label{eq420}
\eeq
They satisfy the axioms (\ref{eq33}) and form the coalgebra. In order to
show the structure of Hopf algebra,
the existence of antipode is necessary. Substituting
(\ref{eq419}) and (\ref{eq420}) into the axiom of the antipode, we
get equations for the antipode of $ a $ and $ b $
\bea
  & a + b\, S(a) = S(a) + S(b)\, a = 0, \nn \\
  & b\, S(b) = S(b)\, b = 1.                                      \label{eq421}
\eea
It is obvious that $ b^{-1} $ is needed to define the antipode.
Incorporating $ b^{-1} $, the antipode of $ a $ and $ b $ is obtained
\beq
   S \left(
   \begin{array}{cc}
   1 & 0 \\
   a & b
   \end{array} \right) =
   \left(
   \begin{array}{cc}
   1 & 0 \\
   a & b
   \end{array} \right)^{-1}
   =
   b^{-1} \left(
   \begin{array}{cc}
   b & 0 \\
   -a & 1
   \end{array} \right),                                         \label{eq422}
\eeq
and this is indeed the inverse matrix of (\ref{eq46}).
We thus complete the
Hopf algebra $ \{a,\, b,\, b^{-1} \} $ adding following properties of
$ b^{-1} $
\bea
   & b\, b^{-1} = b^{-1} \, b = 1, \hspace{1 cm}
     a\, b^{-1} = q^{-1} b^{-1}\,a,                              \label{eq423}
   \\
   & \Delta(b^{-1}) = b^{-1} \otimes b^{-1}, \hspace{1 cm}
     \epsilon(b^{-1}) = 1, \hspace{1 cm}
     S(b^{-1}) = b.                                              \label{eq424}
\eea

  In closing this section, it should be pointed out that
$ \varphi \; : \; F \rightarrow F \otimes {\rm QTAG}, \;
F = \{\td, \tp\} $
is consistent with the coalgebra of (\ref{eq419}) and (\ref{eq420})
\cite{ks}
\bea
   & (id \otimes \Delta) \circ \varphi = (\varphi \otimes id ) \circ \varphi,
   \nn \\
   & (id \otimes \epsilon) \circ \varphi = id.                \label{eq425}
\eea
%
%
\setcounter{equation}{0}
\section{QATG Covariant Differential Calculus}

  In this section, we introduce a two-dimensional plane $ (x^1,\, x^2) $ on
which the QATG acts and develop a differential calculus on the plane.
$ GL_q(n) $ covariant differential calculus on the quantum plane was
already constructed by Wess
and Zumino \cite{wz}. They considered the n-dimensional quantum plane and
require that exterior derivative satisfies the usual properties : the
nilpotency and the Leibnitz rule so that the commutation relations among
coordinates, differentials and derivatives must satisfy various
consistency conditions. The obtained formulae of the differential calculus
are covariant with respect to the action of $ GL_q(n) $.

  According to Ref.\cite{wz}, we wish to construct the differential calculus
covariant with respect to the action of the QATG.
There is, however, a much
difference between Ref.\cite{wz} and the present discussion.
In Ref.\cite{wz}, the quantum plane is firstly defined, and then $ GL_q(n) $
is constructed in order to preserve the commutation relations of the
quantum plane under the action of $ GL_q(n) $.In the present case, we
already have the QATG and we conversely organize a quantum plane whose
commutation relations are preserved under the action of the QATG. Let us
start with the following ansatz for the coordinates $ (x^1,\,x^2) $ and their
differential $ (dx^1,\,dx^2) $
\bea
   & x^1\,x^2 = \kappa\,x^2\,x^1, \nn \\
   & dx^1\,dx^2 = -\lambda\,dx^2\,dx^1,  \label{eq51} \\
   & (dx^i)^2 = 0.                 \nn
\eea
Requiring that (\ref{eq51}) should be preserved under the action of the QATG
\beq
  \left(
  \begin{array}{c}
  \varphi(x^1) \\  \varphi(x^2)
  \end{array} \right)
  =
  \left(
  \begin{array}{cc}
   1 & 0 \\
   a & b
  \end{array} \right)
  \left(
  \begin{array}{c}
  x^1 \\  x^2
  \end{array} \right)_,
  \hspace{0.5 cm}
  \left(
  \begin{array}{c}
  \varphi(dx^1) \\  \varphi(dx^2)
  \end{array} \right)
  =
  \left(
  \begin{array}{cc}
   1 & 0 \\
   a & b
  \end{array} \right)
  \left(
  \begin{array}{c}
  dx^1 \\  dx^2
  \end{array} \right)_,                                   \label{eq52}
\eeq
we can achieve this situation if and only if,
\beq
  \kappa = 1, \hspace{1 cm} \lambda = q^{-1}.              \label{eq53}
\eeq
This defines a new quantum plane, {\em i.e.} the QATG covariant quantum plane,
whose coordinates commute each other but differentials do not. It should
be noted that $ (x^1,\,x^2) $ is also a comodule algebra of the QATG, namely,
the
map $ \varphi $ is consistent with the coproduct and the counit
\bea
   & & (\Delta \otimes id) \circ \varphi = (id \otimes \Delta) \circ \varphi,
   \nn \\
   & & (\epsilon \otimes id ) \circ \varphi = id.         \label{eq54}
\eea

  Now we introduce the derivatives on our quantum plane
\beq
  \partial_i \equiv {\partial \over \partial x^i}, \hspace{1 cm}
  \partial_i \, x^j = \delta_i^j.                          \label{eq55}
\eeq
Since the derivatives should be contravariant, the QATG acts on the derivatives
\beq
  \left(
  \begin{array}{c}
  \varphi(\partial_1) \\ \varphi(\partial_2)
  \end{array} \right)
  =
  \left( S \left(
  \begin{array}{cc}
   1 & 0 \\
   a & b
  \end{array} \right) \right)^T
  =
  \left(
  \begin{array}{cc}
   1 & -ab^{-1} \\
   0 & b^{-1}
  \end{array} \right) \,
  \left(
  \begin{array}{c}
  \partial_1 \\ \partial_2
  \end{array} \right)_,                                 \label{eq56}
\eeq
where $ T $ denotes the transposition. The map (\ref{eq56}) preserves the
relation
(\ref{eq55}) and satisfies the consistency conditions (\ref{eq54}).
The exterior derivative is then defined in the standard way
\beq
  d = dx^i\, \partial_i,                                \label{eq57}
\eeq
where the sum over $ i $ is understood. It is required that $ d $ satisfies
the nilpotency and the Leibnitz rule
\bea
  & d^2 = 0, \nn \\
  & d\,(fg) = (df)\, g + (-1)^f\,f\,dg,                  \label{eq58}
\eea
where $f$ and $g$ are $p$-form and $ (-1)^f $ is $ -1\: (+1) $ if $ f $ is
odd (even) element. If we write (5.1) and (5.3) as
\bea
  & & x^i\, x^j = B^{ij}_{kl}\, x^k\, x^l, \nn \\
  & & dx^i\, dx^j = -C^{ij}_{kl}\, dx^k\, dx^l,          \label{eq59}
\eea
and the commutation relations of the derivatives as operator
\beq
  \partial_i \, \partial_j = F^{lk}_{ji} \,\partial_k \, \partial_l,
                                                         \label{eq510}
\eeq
the requirements of (\ref{eq58}) determine the other commutation relations
as follows :
\bea
  & & x^i\, dx^j = C^{ij}_{kl} \, dx^k\, x^l, \nn \\
  & & \partial_j\, x^i = \delta_j^i + C^{ik}_{jl}\, x^l\, \partial_k,
  \label{eq511} \\
  & & \partial_j\, dx^i = (C^{-1})^{ik}_{jl}\, dx^l\, \partial_k, \nn
\eea
where the matrices $ B$, $ C $ and $ F $ must satisfy the relations \cite{wz}
\bea
  & & B_{12}\, C_{23}\, C_{12} = C_{23}\, C_{12}\, B_{23}, \nn \\
  & & C_{12}\, C_{23}\, C_{12} = C_{23}\, C_{12}\, C_{23}, \label{eq512}  \\
  & & C_{12}\, C_{23}\, F_{12} = F_{23}\, C_{12}\, C_{23}. \nn
\eea

  It is obvious that the eqs.(\ref{eq512}) are satisfied by the following
choice
\beq
  B = F = \ckr^{(1)}, \hspace{1 cm} C = q^{-1} \ckr^{(1)},   \label{eq513}
\eeq
where $ \ckr^{(1)} $ is given in eq.(\ref{eq412}) and eq.(\ref{eq59}) reduces
to eq.(\ref{eq51}). Of course, another solution have we
\beq
  B = F = \ckr^{(2)}, \hspace{1 cm} C = q\, \ckr^{(2)}.   \label{eq514}
\eeq
Eq.(\ref{eq513}) or (\ref{eq514}) completes the differential calculus. It is
not difficult to verify that all relations of the differential calculus
described above are preserved under the action of the QATG, {\em i.e.} the
linear transformation $ \varphi $ given in eqs.(\ref{eq52}) and (\ref{eq56}).

  Now we have obtained two differential calculi on the quantum plane
(\ref{eq51}). As an illustration, we give explicit commutation relations of
the differential calculus in the case of (\ref{eq513}). Denoting the
coordinates by $ (x,\,y) $ instead of $ (x^1,\,x^2) $, the QATG covariant
quantum plane is
\bea
  & & x y = y x,  \nn \\
  & & dx\, dy = -q^{-1}dy\, dx,\label{eq515} \\
  & & (dx)^2 = (dy)^2 = 0. \nn
\eea
The commutation relations between the coordinates and the differentials are
\bea
  & & x \, dx = q^{-1}dx\, x, \nn \\
  & & x \, dy = q^{-1}dy\, x, \nn \\
  & & y \, dx = dx \, y  + (q^{-1}-1)\, dy \, x,  \label{eq516} \\
  & & y \, dy = q^{-1} dy\, y.                   \nn
\eea
We have derivatives which satisfy the relation
\beq
  \partial_x \partial_y = q^{-1} \partial_y \partial_x.      \label{eq517}
\eeq
The commutation relations between the derivatives and the coordinates are
\bea
 & & \partial_x \, x = 1 + q^{-1} x \, \partial_x, \nn \\
 & & \partial_x \, y = y \, \partial_x,            \nn \\
 & & \partial_y \, x = q^{-1} x \, \partial_y,     \label{eq518} \\
 & & \partial_y \, y = 1 F+ (q^{-1}-1)\, x \, \partial_x
      + q^{-1} y \, \partial_y,                    \nn
\eea
and those between the derivatives and the differentials are
\bea
 & & \partial_x\, dx = q\, dx \, \partial_x + (q-1)\, dy, \partial_y,
 \nn \\
 & & \partial_x \, dy = q\, dy \, \partial_x,            \nn \\
 & & \partial_y \, dx = dx \, \partial_y,                 \label{eq519} \\
 & & \partial_y \, dy = q \, dy \, \partial_x.           \nn
\eea
Using the relation (\ref{eq518}), one can calculate the derivative of the
monomial
\[
   \partial_x (x^n\, y^m) = {1 -  q^{-n} \over 1 - q^{-1} } x^{n-1} y^m,
\]
\beq
   \partial_y (x^n\, y^m) = q^{-n}
  {1 -  q^{-n} \over 1 - q^{-1} } x^n y^{m-1}.           \label{eq5199}
\eeq
These formulae are utilized for estimating the derivative of a power series.
In this way, we can calculate any higher derivatives of an arbitrary regular
function of $ x $ and $ y $.

  Finally, we mention another type of the quantum plane which is covariant
under the action of the transposed matrix
\beq
  \left(
  \begin{array}{c}
  \varphi(x^1) \\  \varphi(x^2)
  \end{array} \right)
  =
  \left(
  \begin{array}{cc}
   1 & a \\
   0 & b
  \end{array} \right)
  \left(
  \begin{array}{c}
  x^1 \\  x^2
  \end{array} \right)_.                                 \label{eq520}
\eeq
The covariant quantum plane with respect to (\ref{eq520}) is amount to
\bea
  & & x \, y = q \, y \, x,                    \nn \\
  & & dx \, dy = - dy \, dx,                   \label{eq521}    \\
  & & (dx)^2 = (dy)^2 = 0.                     \nn
\eea
Also in this case, it is possible to construct two differential calculi
on this quantum plane and all their formulae are preserved by the
action (\ref{eq520}). The differential calculi are given by
\beq
   B = F = \ckr'\,^{(1)},  \hspace{1 cm} C = q^{-1} \ckr'\,^{(1)},
                                                       \label{eq522}
\eeq
and
\beq
  B = F = \ckr'\,^{(2)},  \hspace{1 cm}  C = q \ckr'\,^{(2)}.
                                                        \label{eq523}
\eeq
We have, as a result, obtained four types of differential calculus. The
relations among them are recognized through the figure~1.
%
%
\setcounter{equation}{0}
\section{Conclusions}

  In this paper, we discussed quantum deformation of the affine transformation
group and its Lie algebra. The Lie algebra was deformed in the sense of
Drinfeld and Jimbo and it was shown that the QATA possesses the
non-cocommutative
Hopf algebra structure. The tensor product of two adjoint representations
of the QATA has an intertwiner, however the intertwiner was not compatible
with the YBE.
{}From this fact, it was inferred that the QATA is not quasi-triangular. The
deformation, in the sense of Manin, of the affine transformation group
was also accomplished by making use of the adjoint representation.
Manin considered the deformation on the quantum plane whose elements
commute each other in the limit of $ q \rightarrow 1 $, while the
basis of the QATG do not commute even in the limit of
$ q \rightarrow 1 $ because they form the Lie algebra. This is the main
difference between Manin's and our approaches. It was also shown that the QATG
has the Hopf algebra structure.

  The relationship between the QATA and the QATG is not so clear as that of
$ U_q(sl(n)) $ and $ GL_q(n) $. This problem should be discussed elsewhere.
Another interesting problem is a multi-parameter deformation. It will not
be easy since the affine transformation group is too simple to incorporate
many parameters. As was mentioned in \S 1, the affine transformation
corresponds
to the dilatation and the translation in  one-dimensional space. The
generalization to higher dimensional space gives a possibility of the
multi-parameter deformation. If the coordinates
$ x^i \; (i = 1,\,2,\, \cdots,\, N) $ of $N$-dimensional space
commute each other, the generalization is trivial, {\em i.e.} the direct
product of $N$ independent the QATAs and the QATGs. However, if $N$ dimensional
quantum plane is considered, the generalization will be non-trivial.
It would be possible to adopt different deformation parameters for the space
and for the affine transformation groups. In this case, it is natural
to consider the multiparameter quantum plane. For example, multiparameter
deformation of the phase space has discussed in Ref.\cite{fzn}.

  In \S 5, we found the differential calculi which are covariant with
respect to the action of the QATG. To this end, new quantum planes have been
introduced. The differential calculus is a QATG comodule algebra which is
generated by $ \{x^i,\, dx^i,\,\partial_i \} $. All the
commutation relations of the differential calculus are written by using
the $\ckr$-matrix. We have two Hopf algebras of the QATA and the QATG. If they
are
dual each other, the differential calculus discussed here would be
generalized according to the method developed by Schupp et al. \cite{swz}.

  The QATA and the QATG are simple in form and their algebraic structure is
well established. We expect that the QATA and the QATG would be available for
building blocks of other quantum algebras or groups and might play a
crucial role in physics similarly in the case of $ q = 1 $.
%
%
%

%

\newpage

\begin{center}

{\Large Figure Caption}
\end{center}

1. Relationship among the $\ckr$-matrices.
\newpage
%
%
\setlength{\topmargin}{0cm}
\setlength{\oddsidemargin}{0cm}
\setlength{\textheight}{23cm}
\setlength{\textwidth}{18cm}
\setlength{\unitlength}{1 mm}
\begin{picture}(160,170)
\thicklines
\put(20,160){\makebox(0,0){\Large (4.6)}}
\put(140,160){\makebox(0,0){\Large (4.14)}}
\put(20,140){\makebox(0,0){\Large $\check R^{(1)}$}}
\put(140,140){\makebox(0,0){\Large $\check R'\,^{(1)}$}}
\put(20,60){\makebox(0,0){\Large $\check R^{(2)}$}}
\put(140,60){\makebox(0,0){\Large $\check R'\,^{(2)}$}}
\put(80,100){\vector(3,2){40}}
\put(80,100){\vector(-3,2){40}}
\put(80,100){\vector(3,-2){40}}
\put(80,100){\vector(-3,-2){40}}
\put(40,59){\framebox(80,2)}
\put(38,60){\line(2,1){4}}
\put(38,60){\line(2,-1){4}}
\put(122,60){\line(-2,1){4}}
\put(122,60){\line(-2,-1){4}}
\put(40,139){\framebox(80,2)}
\put(38,140){\line(2,1){4}}
\put(38,140){\line(2,-1){4}}
\put(122,140){\line(-2,1){4}}
\put(122,140){\line(-2,-1){4}}
\put(19,75){\dashbox(2,50)}
\put(20,73){\line(1,2){2}}
\put(20,73){\line(-1,2){2}}
\put(20,127){\line(1,-2){2}}
\put(20,127){\line(-1,-2){2}}
\put(139,75){\dashbox(2,50)}
\put(140,73){\line(1,2){2}}
\put(140,73){\line(-1,2){2}}
\put(140,127){\line(1,-2){2}}
\put(140,127){\line(-1,-2){2}}
\put(80,50){\makebox(0,0){\rm \large transpose}}
\put(80,150){\makebox(0,0){\rm \large transpose}}
\put(9,100){\makebox(0,0){\rm \large inverse}}
\put(151,100){\makebox(0,0){\rm \large inverse}}
\put(80,80){\makebox(0,0){\rm \large $ \sigma \check R^T \sigma $
                              with $ q \leftrightarrow q^{-1}$}}
\put(80,7){\makebox(0,0){\rm \huge Figure~1}}
\end{picture}
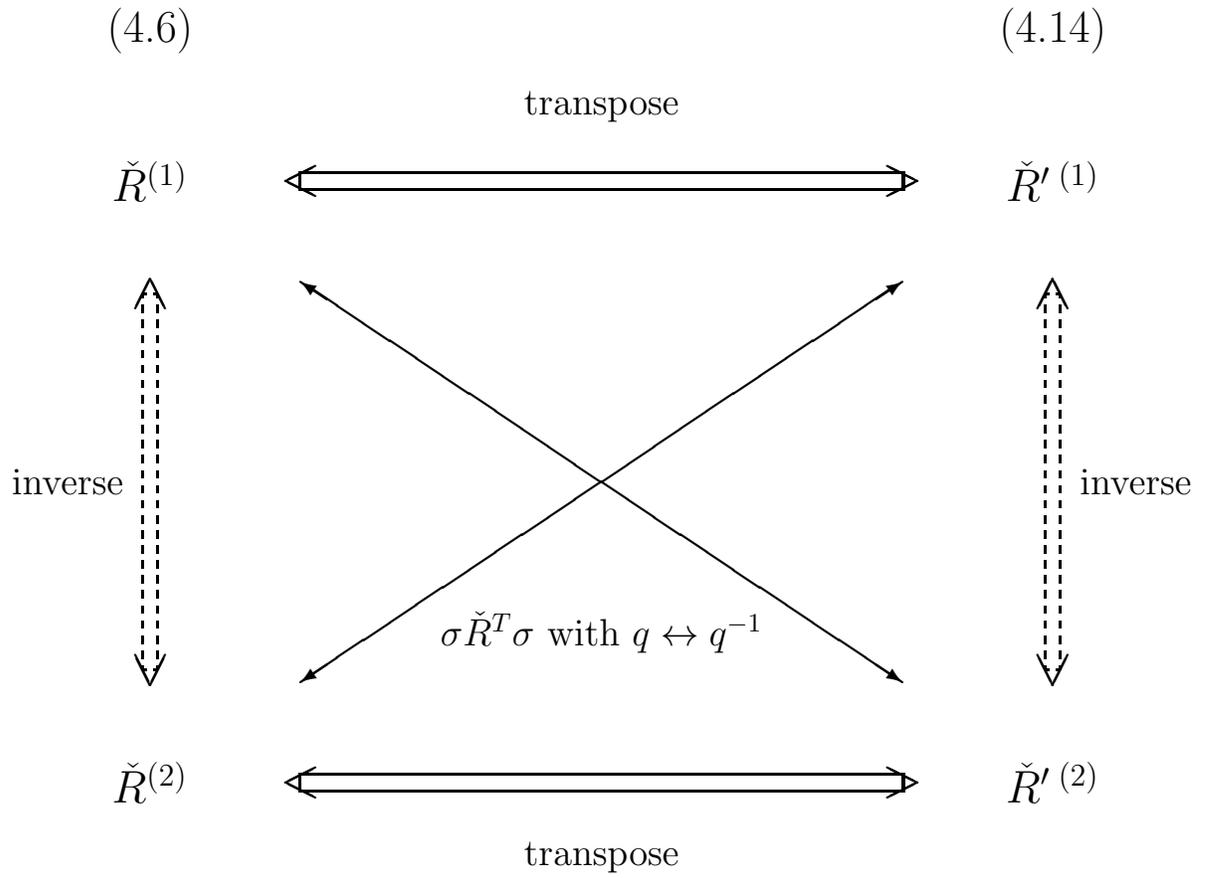
\end{document}